\documentclass[prd,aps,preprintnumbers,nofootinbib,showkeys,showpacs,amsmath,amssymb,superscriptaddress,floatfix]{revtex4}
\usepackage[dviwindo]{graphicx}

\newtheorem{assunptions}{Assumption:}
\newtheorem{corollary}{Corollary:}
\newtheorem{comment}{Comment:}

\begin{document}
\title{Schwarzschild Massive-Point-Particle Problem in Arbitrary Radial Gauge}
\author{Plamen~P.~Fiziev\footnote{ E-mail:\,\, fiziev@phys.uni-sofia.bg  and fiziev@theor.jinr.ru} }
\affiliation{
Sofa University Foundation for Theoretical and Computational Physics and Astrophysics,
5 James Bourchier Boulevard,
Sofia~1164, Bulgaria. \\and\\  BLTF, JINR, Dubna, 141980 Moscow Region, Rusia.}
\begin{abstract}
 We present, for the first time, a correct solution of the Schwarzschild Massive Point (SMP) problem
in {\em arbitrary} radial gauge and formulate the strict mathematical assumptions, which are necessary and sufficient for this.

In GR, there exists a two-parameter family of such exact SMP solutions to the Einstein equations,
which are physically distinguishable from the well-studied one parameter family of vacuum Schwarzschild solutions
related with Schwarzschild Black Holes (SBH).

The obtained here SMP family of solutions is defined by positive bare mass $M_0>0$ and positive Kepler mass $M<M_0$, or, alternatively,
by the standard gravitational radius $\rho_g$ and mass ratio $\varrho=M/M_0 \in (0,1)$.
The metrics of spacetime have an unavoidable jump at the place of a massive point particle.

We also present a proper development of the theory of distribution defined by kernels with a finite jump
which appear in the solution of SMP. The specific properties of these distributions are used for work with SMP.

A series of physical properties of SMP solutions are derived and commented.

Our findings are important for description of Extremely Compact Objects (ECOs)
studied in relation with possible echoes in Gravitational Waves (GW) recently discovered by the LIGO/VIRGO collaboration.


\end{abstract}
%
\sloppy
\newcommand{\lfrac}[2]{{#1}/{#2}}
\newcommand{\sfrac}[2]{{\small \hbox{${\frac {#1} {#2}}$}}}
\newcommand{\ben}{\begin{eqnarray}}
\newcommand{\een}{\end{eqnarray}}
\newcommand{\la}{\label}
\maketitle
%
\section{Introduction}\la{Intro}

\subsection{The Schwarzschild Mass Point Problem}

As early as in February 1916, Karl Schwarzschild published
his famous paper \cite{Schwarzschild} titled "On the Gravitational Field of a Mass Point according to Einstein's Theory".
We dub this subject the Schwarzschild Mass Point (SMP) problem.

It is curious that today the original form of the solution proposed by Schwarzschild is not well known to wide audience,
as well as the correct solution of the very SMP problem.

While at the early stages of development of GR there were no proper mathematical tools for correct solution of the SMP problem,
in modern literature, there exists a incorrect opinion that it is impossible to solve the SMP problem because:

1) the description of a massive point particle needs the use of mass-density distribution described by the Dirac $\delta$-function. This is a correct statement.

2) the implementation of the theory of distributions in GR is impossible because of the nonlinear character of the Einstein Equations.

This is not a completely correct statement, since the Einstein equations are quasi-linear differential equations. The last property gives rise to a nontrivial way
of overcoming the problems with distributions, related with SMP, as shown in the present article.

The wide audience is happy with the form of the {\em vacuum} solution to the Einstein equations, proposed by Hilbert in 1917 \cite{Hilbert1917}
and largely dubbed today as "the Schwarzschild solution of the Einstein equations" in "the Schwarzschild coordinates":
\ben
ds^2=\left(1-{\frac {\rho_g}{\rho}}\right)dt^2- {\frac{d\rho^2}{1-{\frac {\rho_g}{\rho}}}} -\rho^2 d\Omega^2,
\la{Hilbert}
\een
where $d\Omega^2=d\theta^2+\sin\theta^2 d\phi^2$ is the standard metric on the unit 2D sphere
${}^2\!\mathcal{S}$ in 3D Euclidean space ${}^3\!\mathcal{E}$, $\rho$ is the luminosity radius,
and $\rho_g =2GM/c^2$ is the gravitational radius of the spherical body with the Kepler mass M, see detail in \cite{books}.

Further on, as a rule, we use units in which the velocity of light $c=1$.

According to the strict and correct conclusion by Erik J. Weinberg \cite{EWeynberg} about solution \eqref{Hilbert},
"Although we began by seeking the metric generated by a point mass at the origin,
and have indeed found a solution that at large distances is consistent with the Newtonian
potential from such a source, our spacetime has no static point mass inside it."

Indeed, the energy-momentum tensor of the solution \eqref{Hilbert} $T_{\mu\nu}\equiv 0$ vanishes identically  everywhere in spacetime.

The well-known modern interpretation of the vacuum solution \eqref{Hilbert} as the Schwarzschild Black Hole (SBH) is well-studied in detail \cite{books}.

\subsection{Mass Point Problem in the Era of Gravitational Wave Observations}

LIGO detection of Gravitational Waves (GW) is
the most important finding in gravity
after Sir Isaac Newton's discovery of the gravitational field. Indeed,
Sir Isaac Newton discovered the gravitational field attached to bodies.
LIGO discovered the gravitational field detached from bodies and freely spreading in space.
The last being a qualitatively different novel gravitational phenomenon.

Without any doubt, with the LIGO/VIRGO observations and analysis of 11 well-established GW events, included in the First catalog of GW events,
and more than 35 additional candidates of such events from the third scientific run
(see the references at the WEB address \cite{LOGOOPEN}), a new era in fundamental physics  started \cite{Barack2018}.
Finally, the  gravitational astronomy opened a novel window to the Universe and started to give us hitherto unreachable knowledge of the Nature.

These achievements deserve extraordinary careful analysis of all issues, which appear for the first  time  and are accompanied by many uncertainties and unknowns.
Prior ruthless examination of all facts, hypotheses, assumptions, and interpretations, we cannot be sure of what we really see in this newly opened window  to the Universe.

We certainly may refer to the already existing models and theoretical achievements
like General Relativity (GR) well tested in other physical domains
or to a variety of strictly speaking still hypothetical Black Hole (BH) models.

However, the most important thing in the new situation is the newly appearing opportunity to examine experimentally theoretical
assumptions, as well as to look for new developments and unexpected physical phenomena.

The used methods for processing the LIGO data were good enough to discover GW without any doubt
but failed to  recover the most important details needed to establish definitely the right theory
of the observed GW and the physical nature of their sources.

For example, these methods were too crude to recover Quasi Normal Modes (QNMs) as fingerprints of BH,
see, for example, the book by Chandrasechar in \cite{books}, \cite{RW,Starobinskij1973,Z,ChandraDetweiler1975,Kokkotas1999,Nagar05,Fiziev2006,Berti06,Berti09,Fiziev2011,Nigel2016,Testing2019,Barack2018}
and a huge amount of references therein.

The used methods also turned out to be not sufficient to establish the existence
or absence of echoes as fingerprints of Extremely Compact Objects (ECOs)\footnote{
In the last few years the abbreviation "ECO" is used for
Exotic Compact Objects without an event horizon, which are not vacuum BH solutions of gravitational equations.
We use here this abbreviation in a broader sense, including BH in the class of ECOs = {\em Extremely} Compact objects.},
which are not BH \cite{Cardoso16,Mark2017,Cardoso2017,Conklin18,Wang2018,Raposo18,Cardoso2019}.

Between others, GW150914 is the event with the largest signal-to-noise-ratio in the QNMs ringing domain.
None of the methods, recently proposed for finding echoes, have yet found
a significant echo signal in this event \cite{Conklin18} or discarded firmly their existence.

As a result, we still cannot refute firmly many of the alternative theoretical
explanations for the sources of  GW events observed by the LIGO/VIRGO collaboration.

For recent reviews on the above problems, as well as on a series of other ones,
see the talk by Alan Weinstein at the First LIGO Open Data Workshop 2018 \cite{LOGOOPEN}
and a large review papers on the general situation in gravity \cite{Barack2018,Cardoso2019}.

The most recent observation of the shadow of the ECO
at the center of the elliptic galaxy M87 proves the extremely compact character of the central object \cite{EHT},
but it does not give any indication of the presence or absence of an event horizon of this object.
Thus, we have no correct proof that the central object in M87 is a BH.

The main hypothesis during the analysis of the observed GW events is that behind them is
a merger of Binary Black Holes (BBH), Binary Neitron Stars (BNS) or Black Hole - Neutron Star Binaries (BHNSB).
This process includes three phases: Inspiral, Merger and Ringing (IMR).

In numerical calculations, during the inspiral phase the objects in binaries of all types are considered
as point particles with given Kepler masses $M_{i=1,2}$ and coordinates ${\bf x}_i(t)$ described by the point-mass energy-momentum tensor
\ben
T^{\mu\nu}=\sum_i {\frac {M_i}{\sqrt{-{}^4\! g}}} {\frac {dx^\mu}{ds}}{\frac {dx^\nu}{dt}} \,{}^3\!\delta \left(  {\bf r}-{\bf r}_i(t) \right),
\la{EM_MPs}
\een
see for example \cite{books,Maggiore08,Jafari19}, and the references therein.

It becomes obvious that the first step to the correct mathematical treatment of such problems with distributional energy-momentum tensors
is a rigorous solution of the SMP problem.

This turns to be not a simple issue. The present paper is  further development of our previous study of SMP \cite{Fiziev2019}.

Here we consider for the first time the SMP in arbitrary radial gauge and clarify the mathematical assumptions, needed for this purpose, as well as their physical consequences.

\section{Gauges in General Relativity}\la{Gauges}

In GR, the Einstein equations:
\ben R^\mu_\nu - {\tfrac 1 2}R \,\delta^\mu_\nu = \kappa T^\mu_\nu,\quad \mu,\nu=0,1,2,3;
\la{Einst}
\een
determine the solution of a given
physical problem up to four arbitrary functions, i.e., up to a
choice of coordinates. This reflects the well-known fact that GR
is a specific kind of gauge theory. In Eqs.\eqref{Einst} $\kappa=8\pi G/c^4$
denotes the Einstein constant.

According to the Landau-Lifshitz book in \cite{books},
the fixing of the gauge in GR in a {\em holonomic}
frame is represented by a proper choice of the quantities
\ben
\bar\Gamma_\mu = -{{1}\over{\sqrt{-{}^4\!g}}}g_{\mu\nu}
\partial_\lambda \left( \sqrt{ -{}^4\! g} g^{\lambda\nu} \right),
\la{Gammas}
\een
which emerge when one expresses the 4D d'Alembert operator in the form
$g^{\mu\nu}\nabla_\mu\nabla_\nu=g^{\mu\nu}
\left(\partial_\mu\partial_\nu-\bar\Gamma_\mu\partial_\nu\right)$.

We shall call the change of the gauge fixing expressions (\ref{Gammas}),
{\em without} any preliminary conditions on the analytical behavior
of the used functions,  gauge transformations in a {\em broad} sense.
This way we essentially expand the class of
admissible gauge transformations in GR \cite{Fiziev2019}.

\section{Static Spherically Symmetric Space-Times with a Single Point Source of
Gravity at rest}\la{SPS}
For a clear presentation of the ideas and results we remind some basic relations between the Special Theory of Relativity (SR) and GR \cite{books}.

In SR, a basic assumption is that one has to work with the 4D Minkowski spacetime ${}^{(1,3)}\!\mathcal{E}\{x,\eta_{\mu\nu}(x)\}$ 
equipped with some {local} coordinates $x=\{ x^\mu \}$ and corresponding flat metric $\eta_{\mu\nu}(x)$ of signature $\{+,-,-,-\}$,
which annuls the Riemann curvature tensor $R_{\mu\nu,\lambda}{}^\kappa = 0$. 
Transition to new (curvilinear) coordinates $y=f(x)$ can be considered as a change of gauge and the corresponding gauge functions
\ben
\bar\Gamma_\mu=-{{1}\over{\sqrt{-{}^4\!\eta}}}\eta_{\mu\nu}
\partial_\lambda\left(\sqrt{-{}^4\!\eta}\eta^{\lambda\nu}\right).
\la{GammasSR}
\een
For example, in the Cartesian coordinates $\eta_{\mu\nu}=\text{diag}\{1,-1,-1,-1\}$ and $\bar\Gamma_\mu=0$, 
while using the spherical 3D coordinates ${\bf r}=\{x^1,x^2,x^3\}=\{r\sin\theta\cos\phi, r\sin\theta\sin\phi, r\cos\theta\}$ we obtain
the 4D interval in the Minkowski space-time in the form
\ben
ds^2 = dt^2-dr^2 -r^2 d\Omega^2, \quad \text{and}
\la{dsMinkowski}
\een
\ben
\bar\Gamma_t=0,\quad  \bar\Gamma_r=-{\frac 2 r},\quad \bar\Gamma_\theta= - \cot \theta,\quad \bar\Gamma_\phi =0.
\la{Gammas_Minkowski}
\een

In GR, the basic assumption is that geometry of the 4D spacetime is described by the pseudo Riemannian manifold  
${}^{(1,3)}\!\mathcal{M}\{x, g_{\mu\nu}(x)\}$ of signature $\{+,-,-,-\}$, 
equipped with some coordinates $x=\{ x^\mu \}$ and metric $g_{\mu\nu}(x)$ that solves the Einstein equations \eqref{Einst}.

The tangent space ${}^{(1,3)}T_x\mathcal{M}\{x, g_{\mu\nu}(x)\}$  above each point $x \in {}^{(1,3)}\!\mathcal{M}\{x, g_{\mu\nu}(x)\}$ is 
 the Minkowski space-time of SR and it defines the local frame in which our laboratory experiments are performed 
 and all local physical laws are established. The connection of the tangent spaces at different 
 spacetime points $x$ is defined by the corresponding Levi-Chivita connection \cite{books}.

We can use the same {\em local} coordinates in ${}^{(1,3)}T_x\mathcal{M}\{x, g_{\mu\nu}(x)\}$ and in ${}^{(1,3)}\!\mathcal{M}\{x, g_{\mu\nu}(x)\}$ 
in some vicinity of the point $x$. In some specific cases,  part of these common local coordinates may be global.

In the GR-spacetime ${}^{(1,3)}\!\mathcal{M}\{x, g_{\mu\nu}(x)\}$  with a massive point in it, 
there exists unambiguous choice of a global time $t$, due to the requirement to use a static metric 
in the rest frame of the point particle. We assume that the massive point is placed at the geometrical point $r=0$ 
that defines the coordinate origin.

This yields a familiar form of the space-time interval \cite{books}:
\ben
ds^2=g_{tt}(r)\,dt^2+g_{rr}(r)\,dr^2-\rho(r)^2d\Omega^2
\la{ds0}
\een
where the functions $g_{tt}(r)>0,\,g_{rr}(r)<0,\,\rho(r)>0$ have to be obtained from 
the Einstein equations \eqref{Einst} under the corresponding gauge conditions \eqref{Gammas}
and corresponding boundary conditions.

Note that in the frame of free falling clocks,
if $\rho_{fixed}$ is some arbitrary fixed value of the luminosity distance $\rho$, 
at which the clocks start to fall with zero velocity, in the Schwarzschild vacuum space-time the expression
$$\left(\rho_2-\rho_1\right)/\left( 1-2M/\rho_{fixed}\right)^{1/2}$$
measures the 3D geometrical distance between the geometrical points
$2$ and $1$ on a radial geodesic line \cite{GH}.
Nevertheless, even in this frame the absolute value of the variable
$\rho$ remains not fixed by the 3D distance measurements.

In the case at hand, the form of three of the gauge fixing
functions  \eqref{Gammas} $\bar\Gamma_t, \bar\Gamma_\theta$, and $\bar\Gamma_\phi$ is the same as in Eqs.\eqref{Gammas_Minkowski}.
The only exception is $\bar\Gamma_r=-{\frac {d\bar\varphi}{dr}}$, where
\ben
\bar\varphi(r)=
\ln\left({\frac {\rho^2} {a^2}} \sqrt{{\frac {g_{tt}} {-g_{rr}}}}\right)
\la{barGamma_r}
\een
with some constant $a$ with dimension of length.

The function $\rho(r)$ must be fixed in a rather arbitrary way.
We refer to the freedom of choice of the function $\rho(r)$ as "{\em radial-gauge
freedom}", and to the choice of the $\rho(r)$ function as
"{\em radial-gauge fixing}" in a {\em broad} sense \cite{Fiziev2019}.

After one chooses the function $\rho(r)$, the Einstein equations \eqref{Einst} define under 
the corresponding boundary conditions the unknown functions $g_{tt}(r)$ and $g_{rr}(r)$ unambiguously.

In the present article, we will not use more general gauge transformations in a broad sense than
the radial-gauge ones defined, for example,  by the choice of the function $\rho(r)$, or in some equivalent way.

\subsection{Three Dimentional Riemannian Geometry of the Point Particle Problem}\la{3DRG}

The single point particle with proper rest bare mass $M_0>0$
can be treated as a 3D entity. Its proper frame of reference is
most suitable for description of the {\em static} space-time with this
single particle in it.

We prefer to reduce the problem of single point source
of gravity in GR to a 1D mathematical problem, considering the
dependence of the corresponding functions on the only essential
variable -- the radial variable $r$. This can be achieved in
the following way.

The spherical symmetry of the 3D space reflects adequately
the non-rotating-point character of the source of gravity.

The spherically symmetric 3D Riemannian space
${}^3\!\mathcal{M}\{-g_{mn}({\bf r})\} \subset {}^{(1,3)}\!\mathcal{M}\{g_{\mu\nu}(x)\}$, $m,n=1,2,3$
can be described using the standard spherical coordinates
$r\in[0,\infty),\,\theta\in[0,\pi),\phi\in[0,2\pi)$
in some auxiliary Euclidean space ${}^3\!\mathcal {E}$\footnote{
The natural idea to use such auxiliary Euclidean space in
the problem under consideration  can be found already in \cite{Schwarzschild},
but Schwarzschild applied different coordinates in it and did not relate this 
Euclidean space with tangent space at the position of the point particle in GR.}.
Then, as usual
and an Euclidean squared 3D infinitesimal distance
$$dl^2=dr^2+r^2 d\Omega^2.$$

The role of GR is that the Einstein Eqs.\eqref{Einst} transform the Euclidean space
${}^3\!\mathcal{E}\{-\eta_{mn}(x^1,x^2,x^3)\} $ to some curved Riemannian one
${}^3\!\mathcal{M}\{-g_{mn}((x^1,x^2,x^3))$ with a squared 3D infinitesimal distance

\ben
dl^2=- g_{rr}(r)\,dr^2+\rho(r)^2 d\Omega^2
\la{dl0}
\een
with beforehand unknown functions $g_{rr}(r)<0$ and $\rho(r)>0$, 
which have to be obtained from the Einstein equations \eqref{Einst} 
corresponding to the initial and boundary conditions and making certain choice of the gauge functions \eqref{Gammas}.

The physical and geometrical meaning of the radial coordinate $r$ in space ${}^3\!\mathcal{M}\{-g_{mn}({\bf r })\}$
is not defined by the spherical symmetry of the problem and is
unknown \textit{a priori} \cite{Eddington}.

The only clear thing
is that its value $r=0$ corresponds by construction to the center of the symmetry,
where one must place the physical source of the gravitational
field\footnote{In the present article, we assume that in the Universe
other sources of the gravitational field outside the center of
symmetry do not exist.}.

All these facts permit us to identify the  Euclidean space
${}^3\!\mathcal{E}$ with the tangent space above the place of the point source of gravity, which clarifies
its geometric meaning from a point of view of modern differential geometry.

Now the physical meaning of the space ${}^3\!\mathcal{E}\{-\eta_{mn}\}$ becomes clear.
This Euclidean space is precisely the tangent 3D space to
${}^3\!\mathcal{M}\{-g_{mn}\}$ at the place of the massive point where we are performing all our measurements
with our laboratory equipment staying in the rest reference frame of the point particle.

The space ${}^3\!\mathcal{E}\{-\eta_{mn}\}$ is a subspace of the tangent space to the
4D basic manifold ${}^{(1,3)}\!\mathcal{M}\{g_{\mu\nu}(x)\}$, briefly described in the previous Section \ref{SPS}.

GR permits us to relate the measurements in this reference system with measurements
in other ones above other points in the base space ${}^{(1,3)}\!\mathcal{M}\{g_{\mu\nu}\}$ 
using the Levi-Chevia connection defined by metric $g_{\mu\nu}$ \cite{books}.

The radial 3D geometrical distance from the point source at the center of the symmetry is
\ben
l(r)=\int_0^r \sqrt{-g_{rr}(r)} dr
\la{ldistance}
\een

The quantity $\rho$ in Eq.\eqref{dl0} has a clear
geometrical and physical meaning: It is well known that $\rho$
defines the area
\ben
A=4\pi\rho^2
\la{Area}
\een
of the centered at $r=0$ sphere
with the {\em luminosity} radius $\rho$ and the length of a big circle on it
\ben
l_\rho=2\pi\rho.
\la{lcircle}
\een
Relations \eqref{Area} and \eqref{lcircle} only resemble the Euclidean geometry ones
since the variable $\rho$ does not measure radial geometrical distances in the curved space ${}^{(1,3)}\!\mathcal{M}\{-g_{mn}\}$.

One can refer to the quantity  $\rho$ as an {\em ``area
radius''},
or as an optical
{\em "luminosity distance``}, because the luminosity of
distant physical objects is reciprocal to $A$, see the book by J. L. Synge in \cite{books}.

This consideration shows that the area radius $\rho$ is an invariant quantity
and its value does not depend on the choice of coordinates in the 3D Riemannian space ${}^3\!\mathcal{M}\{-g_{mn}\}$.
Hence, the area radius $\rho$ itself can be used as a preferable radial coordinate in the SMP problem \cite{Hilbert1917}.

We shall also mention that in 3D geometry with metric \eqref{dl0}
the 3D volume of a ball with the r-radius $r$ and the center at the origin is
\ben
{}^3V(r)=4\pi \int_0^r \sqrt{-g_{rr}(r)}\, \rho(r) r dr.
\la{3DV}
\een

\subsection{Normal Field-Variables for Gravitational Field of a Massive Point Particle in GR}\la{NormalCoordinates}

An obstacle in describing the gravitational field of a point
source at the initial stage of development of GR was the absence
of an adequate mathematical formalism.

Even after the development
of the correct theory of mathematical distributions \cite{Gelfand}
there still exists an opinion that this theory is inapplicable to
GR because of the nonlinear character of the Einstein equations
(\ref{Einst}).

For example, the author of \cite{YB}
emphasizes that "the Einstein equations, being non-linear, are
defined essentially, only within framework of functions. The
functionals, introduced in ... physics and mathematics (Dirac's
$\delta$-function, "weak" solutions of partial differential
equations, distributions of Schwartz) are suitable only for linear
problems, since their product is not, in general, defined."

In a recent article \cite{GT}, the authors have considered
singular lines and surfaces using mathematical distributions.

Those authors stressed that "there is apparently no viable treatment
of point particles as concentrated sources in GR". See also
\cite{Exact} and references therein.

Here we propose a novel approach to this problem.
It is based on a specific choice of the field variables of the gravitational field.

Let us represent the metric (\ref{ds0}) of the problem at
hand in the specific form:
\ben
ds^2=e^{2\varphi_1}dt^2 -e^{\!-2\varphi_1+4\varphi_2-2\bar\varphi}dr^2\!-
a^2e^{-2\varphi_1+2\varphi_2}d\Omega^2\!,\,\,\,
\la{nc}
\een
and $\varphi_1(x)$,
$\varphi_2(x)$ and $\bar\varphi(x)$ are unknown functions of the
dimensionless variable $x=r/a$. The constant $a$ is the unit of
the luminosity distance $$\rho=a e^{-\varphi_1+\varphi_2}.$$

It is easy to obtain that $\sqrt{-{}^4\!g}=a^2 \sin(\theta)e^{\!-2\varphi_1+4\varphi_2-\bar\varphi}$, the $4D$ scalar curvature relation
\ben
{\frac {a^2} 2}\,{}^4\!R=\left( (\varphi_1^\prime)^2+\varphi_1^\prime \bar\varphi^\prime - \varphi_1^{\prime\prime}
- (\varphi_2^\prime)^2+ 2\varphi_2^\prime \bar\varphi^\prime - 2\varphi_2^{\prime\prime})\right) e^{\!2\varphi_1-4\varphi_2+2\bar\varphi} -  e^{2\varphi_1-2\varphi_2},
\la{R}
\een
and the relation $a\bar\Gamma_r=\bar\Gamma_x=-\bar\varphi^\prime$. Hence, the function $\bar\varphi(x)$
defines the radial gauge of the SMP problem in normal field variables.

\subsection{The Schwarzschild Massive-Point-Particle Problem as 1D Variational Problem}\la{VP}

Taking the integration in the intervals $t\in [t_1,t_2]$ $r\in [r_1,r_2]\ni 0$, $\theta\in [0,\pi]$, $\phi\in [0,2\pi]$, 
one obtains for the total action of the system of GR gravitational field and a point mass with bare mass $M_0$\footnote{The bare mass $M_0>0$ is defined, as usual, 
as a mass of the physical system with gravitational interaction turned off. 
See, for example, the Landau-Lifshitz book in \cite{books}. 
In the case of a single massive point particle at rest, the energy of the heat motion is also excluded from bare mass. 
In the neutron star physics the term "baryon" mass is used often for the bare mass. 
This is not precisely the case, since the bare mass can be associated not only with baryons.}  
being at rest, i.e. with $3D$ velocity  ${\bf v}_{M_0}=0$ at the origin ${\bf r}_{M_0}=0$, one obtains
$$\mathcal{A}_{total}=-{\frac {1}{16\pi G}}\int d^{4}x \sqrt{-{}^4\!g}\,{}^4\!R- M_0 \int ds=
-{\tfrac {a \Delta t}{G}}\,\alpha +\left({\tfrac {a \Delta t}{2G}}\left(\varphi_1^\prime -2\varphi_2^\prime\right)e^{\bar\varphi}\right)\Big|_{x_1}^{x_2}$$
where
\ben
\alpha=\int_{x_1}^{x_2}dx\left({\tfrac 1 2}\left((\varphi_1^\prime)^2e^{\bar\varphi} + {\tfrac {2GM_0}{a}} e^{\varphi_1(x)}\delta(x)\right)  -{\tfrac 1 2} \left((\varphi_2^\prime)^2 e^{\bar\varphi} - e^{2\varphi_2 -\bar\varphi}\right) \right).
\la{alpha}
\een

 The form of the functional \eqref{alpha} justifies the name "normal field variables" for the $\varphi_{1,2}(x)$ and shows that this functional does not depend on the derivatives of the function $\bar\varphi$.

 The massive point particle is described by the standard Dirac function $\delta(x)$ in the functional $\alpha$. The equality
 $e^{\varphi_1(x)}\delta(x)=e^{\varphi_1(0)}\delta(x)$ can be used if $e^{\varphi_1(0)}$ is a well-defined quantity.
 This way we see that the dependence of the functional on the field $\varphi_1(x)$ deviates from 
 the quadratic one only at the place of the massive point $x=0$, i.e. this dependence is quasi-quadratic.

Further on we accept the following
\begin{assunptions}
The value $\varphi_1(0)=\ln\varrho$ is a well-defined and finite real quantity, i.e.  at the place $x=0$ of the point massive particle the value $\sqrt{g_{tt}(0)}=\varrho >0$ is well defined and finite.
\la{Ass1}
\end{assunptions}

Note that the value $\varrho=0$ must be excluded from our consideration. Indeed, if $\varrho=0$, then the Dirac $\delta$-term in the functional \eqref{alpha} disappears and we will turn to a vacuum case.

Then we can define a new quantity\footnote{As we shall see below, the mass $M$ is precisely the Kepler mass of the massive point source 
in the Newton gravity. The mass $M$ is a measurable and finite physical quantity. This is the physical justification of Assumption 1.}
\ben
M = M_0 e^{\varphi_1(0)} = \varrho M_0 \neq 0, \quad \Leftrightarrow \quad \varrho=M/M_0 \neq 0.
\la{M}
\een

Now it becomes clear that the natural choice of the constant $a$ is
\ben
a=GM,   \quad \Leftrightarrow \quad  \rho= GM e^{-\varphi_1+\varphi_2}.
\la{a}
\een

 In the present paper, we do not need to take into account the correction of the Hilbert-Einstein action with 
 the well-known Gibbons-Hawking-York counter-term. 
 Considering 1D variational approach to the Schwarzschild mass point, we can use the functional \eqref{alpha} 
 ignoring all boundary terms \footnote{The same results can be derived directly from the Einstein  
 equations\eqref{Einst} \cite{Fiziev2019} using the corresponding stress-energy tensor for a massive point particle \cite{books}.}.

 Then, the variation of the functional \eqref{alpha} with respect to the fields $\varphi_{1,2}$ gives the second order differential equations 
\begin{subequations} \la{odes:ab}
\ben
\left(\varphi_1^\prime e^{\bar\varphi}\right)^\prime&=&\delta(x),\la{odes:a}\\
\left(\varphi_2^\prime e^{\bar\varphi}\right)^\prime&=&e^{2\varphi_2-\bar\varphi} .\la{odes:b}
\een
\end{subequations}

The variation of the functional \eqref{alpha} with respect to the fields $\bar\varphi$ gives the constraint (weak equation)\footnote{Here and further on 
the symbol $\stackrel{w}{=}$ denotes the equalities in a weak sense, i.e. equalities which are valid only when the constraint \eqref{w} is satisfied.}
\ben
\left(\varphi_1^\prime\right)^2 -\left(\varphi_2^\prime\right)^2 +e^{2\varphi_2-2\bar\varphi} \stackrel{w}{=} 0.
\la{w}
\een

According to the general theory of constrained systems \cite{Dirac_Bargman}, the weak relation\eqref{w} is of the first kind and must be applied only after we find the solutions of the system \eqref{odes:ab} and impose on them all needed additional conditions. Since in the problem under consideration, Eq. \eqref{w} is the only constraint of the first kind,
we have no any constraints of the second kind, which can be obtained by constructing the Dirac brackets between different constraints of the first kind.

\subsection{Solutions of the system \eqref{odes:ab} and additional physical assumptions imposed on them}\la{Solutions}

\subsubsection{General solutions of the system \eqref{odes:a}}\la{GeneralSolution}

Further on we accept the following
\begin{assunptions}
The function $\bar\varphi(x) \in \mathcal{C}^1\{(x_{min},x_{max})/0\}$ in some open interval $(x_{min},x_{max}) \ni 0$, 
i.e. it is at least one time differentiable function with the continuous first derivative in this interval. 
The jump $j_{\bar\varphi}(0)=\bar\varphi(+0)-\bar\varphi(-0)$ is a finite quantity: $|j_{\bar\varphi}(0)| < \infty$.
\la{Ass2}
\end{assunptions}
and develop for the first time the general theory of solution of the Schwarzschild massive-point-particle problem under arbitrary gauge of that class. This way we guarantee the existence of left and right derivatives $\bar\varphi(-0)$ and $\bar\varphi(+0)$.

The solution of the same problem under the simplest regular radial gauge $\bar\varphi_{regular}(x) \equiv 0$ 
was considered in the previous paper \cite{Fiziev2019}. There  were missed some important details and we will fill this gap in the present paper.

The general solution of Eq. \eqref{odes:a} can be obtained in two steps:

1) Taking $\int_{-0}^{+0} dx$ from its left and right sides one obtains
\ben
\varphi_1^\prime(+0)e^{\bar\varphi(+0)}=
\left\{
\begin{aligned}
1 +  \varphi_1^\prime(-0)e^{\bar\varphi(-0)}: \quad x\geq 0,\\
\varphi_1^\prime(-0)e^{\bar\varphi(-0)} : \quad x<0,
\end{aligned}
\right.
\la{varphi_1_prime}
\een
which shows that at the point $x=0$ the function $\varphi_1^\prime(x)e^{\bar\varphi(x)}$ has a finite jump
\ben
j_{\varphi_1^\prime e^{\bar\varphi}}(0)=\varphi_1^\prime(+0)e^{\bar\varphi(+0)}-\varphi_1^\prime(-0)e^{\bar\varphi(-0)}=1.
\la{jphi1}
\een

2) Using the second integration  $\int_{0}^x dx$ and Eq.\eqref{jphi1}, one obtains from Eq.\eqref{varphi_1_prime}
\ben
\varphi_1(x)=
\left\{
\begin{aligned}
\varphi_1^\prime(+0)e^{\bar\varphi(+0)}\Phi(x)+ \varphi_1(0): \quad x \geq 0, \\
\varphi_1^\prime(-0)e^{\bar\varphi(-0)}\Phi(x)+ \varphi_1(0): \quad x<0,
\end{aligned}
\right.
\la{varphi_1}
\een
where
\ben
\Phi(x)=\int_0^x e^{-\bar\varphi(x)}dx, \quad \Rightarrow  \quad \Phi(0)=0
\la{Phi}
\een
is well defined due to Assumption \ref{Ass2}.

Obviously, $\varphi_1(\pm 0)=\varphi_1(0)$, i.e., the function $\varphi_1(x)\in \mathcal{C}^0$ is continuous in some vicinity of the point $x=0$, despite the finite jump \eqref{jphi1} of its derivative.

\subsubsection{General solutions of the system \eqref{odes:b}}\la{Solution_b}

The general solution of Eq. \eqref{odes:b} can also be obtained in two steps:

1) Taking $\int_{0}^x dx$ from its left and right sides after multiplying them by factor $\varphi_2^\prime e^{\bar\varphi}$, one obtains the first integral
\ben
I_2(\varphi_2,\varphi_2^\prime;\bar\varphi)= \left(\varphi_2^\prime e^{\bar\varphi}\right)^2- e^{2\varphi_2}= C^2=const \quad - \quad \text{on the solutions}.
\la{I2}
\een
In general, the constant $C^2$ may be positive, negative, or zero, which defines three different types of solutions.

2) Solving the quadratic Eq.\eqref{I2} with respect to $\varphi_2^\prime$ and integrating once more the result, one obtains the general solutions of Eq. \eqref{odes:b} in the form
\ben
e^{\varphi_2^\pm(x)}=
\left\{
\begin{aligned}
{\frac {C}{\sinh\left( \text{arcsinh}\left(Ce^{-\varphi_2^\pm(0)}\right)\mp C \Phi(x)\right)}}& : \quad |C| \neq 0,\\
{\frac {1}{e^{-\varphi_2^\pm(0)}\pm\Phi(x)}}& : \quad C=0.
\end{aligned}
\right.
\la{varphi_2pm}
\een
 Obviously, the functions $\varphi_2^\pm(x)$ are meromorphic in the complex plain $\mathbb{C}_x$ and increase infinitely 
 at zeros of the corresponding dominators in Eq.\eqref{varphi_2pm}.

 As seen, it is not hard to find the general solution of Eqs.\eqref{odes:ab} for a massive point particle. 
 In this specific case, the nonlinear character of the Einstein equations \eqref{Einst} yields 
 a nontrivial dependence of these solutions on boundary conditions, which will be justified in the next subsection.

\subsubsection{Additional physical assumptions imposed on solutions of the system \eqref{odes:ab}}\la{Assumptions}

1. The physical interval of the radial variable.
\vskip .3truecm

It was already mention in \cite{Schwarzschild} that the only physically admissible singularity in the physical 
domain of spacetime of a massive point particle is at the place of the very particle. 
All other singularities, which may appear in the course of the analysis of the problem, must belong to the nonphysical domain of variables, 
which in general, can be considered as a complex one. 
In our 1D approach, the problem is much simpler, and at this particular point we can restrict our consideration to 
the real axes $\mathbb{R}^{(1)}_x \subset \mathbb{C}^{(1)}_x$.

 Due to the factor $e^{\varphi_2(x)}$, the luminosity radius $\rho(x)$ \eqref{a} becomes infinite at some point
 $x_\infty=x_{max}$, where $x_{max}$ was introduced in Assumption \ref{Ass2}.
 According to Eq.\eqref{varphi_2pm}, at this point  $\Phi(x_\infty)=\pm e^{-\varphi_2^\pm(0)}$.

\begin{assunptions}
The physical domain of the solutions is  $x \in [0, x_\infty )$ for $x_\infty>0$.
\end{assunptions}

\begin{corollary}
Then Eq.\eqref{Phi} shows that the physical general solution in Eq.\eqref{varphi_2pm} is $\varphi_2^{+}(x)$, i.e.,
\ben
e^{\varphi_2(x)}=
\left\{
\begin{aligned}
{\frac {C}{\sinh\left(C\left(\Phi(x_\infty)- \Phi(x)\right)\right)}}& : \quad |C| \neq 0,\\
{\frac {1}{\Phi(x_\infty)-\Phi(x)}}& : \quad C=0.
\end{aligned}
\right.
\la{varphi_2}
\een
\end{corollary}

\begin{corollary}
In addition we have
\ben
\Phi(x_\infty)=
\left\{
\begin{aligned}
C^{-1}\text{\rm arcsinh} \left(Ce^{-\varphi_2(0)}\right)>0 : \quad |C| \neq 0, \\
e^{-\varphi_2(0)} > 0 : \quad |C| = 0,
\end{aligned}
\right.
\la{Phi_infty}
\een
\ben
\rho(x) \geq 0, \quad \text{for}\quad x \in [0, x_\infty ), \quad \text{and}\quad \rho(x_\infty-0)=+\infty.
\la{Phi_infty}
\een
\end{corollary}

From a physical point of view the above results are quite satisfactory;
nevertheless, it is somewhat strange to have a finite interval $r\in [0,MGx_\infty)$
for the radial variable $r$ in the Euclidean space ${}^3\!\mathcal{E}$, equipped with standard nspherical coordinates $r,\theta,\phi$.

We can easily overcome this problem using fractional linear transformation
\ben
x=\left({\frac 1 y}+{\frac 1 {x_\infty}}\right)^{-1} \in [0,x_\infty) \quad \leftrightarrows \quad
y=\left({\frac 1 x}-{\frac 1 {x_\infty}}\right)^{-1} \in [0,\infty).
\la{x-y}
\een

Now the new radial variable $r=MGy$  runs in the usual physical domain $r\in [0,\infty)$\footnote{For simplicity we will not introduce a new notation for the radial variable $r=MGy$ in the Euclidean space ${}^3\!\mathcal{E}_{r,\theta,\phi}$. Instead, in the text we will point explicitly which one we mean: $r=MGy$ or $r=MGx$}.

To some extent, the action of this transformation resembles the one of the well-known Penrouse conformal transformation \cite{books}.

From an analytical point of view the transformation \eqref{x-y} is much better:
It does not change the singularities of the solutions in the whole complex domain $\mathbb{C}^{(1)}_x$.
It only changes the place of the singular point $x_\infty$,  preserving the place of the singular point $x=0$.
Thus, the transformation \eqref{x-y} is indeed a pure "change of the labels" in the Euclidean space ${}^3\mathcal{E}_{r,\theta,\phi}$.

\begin{assunptions}[Einstein]
Asymptotic flatness: Infinitely far from the point mass, 
the four interval \eqref{ds0} must tend to the Minkowski one \eqref{dsMinkowski}  \cite{Einstein1915, {Einstein1916}}.
\end{assunptions}

This assumption was used also in Schwarzschild's paper \cite{Schwarzschild} with reference to the Einstein's one \cite{Einstein1915}.

Taking into account relations \eqref{x-y}, it is not hard to derive the following
\begin{corollary}
To have asymptotically flat metric \eqref{nc} of a massive point, the following limits must take place:
\begin{subequations}\la{Asymtotic:abc}
\ben
&\lim_{x \to x_\infty-0}&\left( e^{\varphi_1(x)}\right)=1,\quad \rightleftarrows\quad \varphi_1(x_\infty)=0,
\la{Asymtotic:a}\\
&\lim_{x \to x_\infty-0}&\left( e^{\varphi_2(x)} \left({\frac {1}{x}} - {\frac {1}{x_\infty}}\right)\right)=
\left(x_\infty^2 e^{-\bar\varphi(x_\infty)}\right)^{-1},
\la{Asymtotic:b}\\
&\lim_{x \to x_\infty-0}&\left( e^{-\bar\varphi(x)}x x_\infty \right)=x_\infty^2 e^{-\bar\varphi(x_\infty)}.
\la{Asymtotic:c}
\een
\end{subequations}
\end{corollary}

From asymptotic conditions \eqref{Asymtotic:a} one easily obtains the following relations:
\begin{subequations}\la{AFC:abc}
\ben
&\Phi(x_\infty) = \ln{\frac 1 \varrho} \in (0,\infty),&
\la{AFC:a}\\
&\varrho \in (0,1),&\la{AFC:b}\\
&\varphi_1^\prime(+0)e^{\bar\varphi(+0)}\Phi(x_\infty)+\varphi_1(0)=0.&
\la{AFC:c}
\een
\end{subequations}

As a result we obtain:

i) Equation \eqref{AFC:a} defines $x_\infty>0$ for a given gauge function $\varphi(x)$ in the form      $\varrho=\exp\left(-\int_0^{x_\infty} e^{-\bar\varphi(x)}dx\right) \in (0,1)$.

ii) Equation \eqref{AFC:b} presents the final justification of the physically admissible domain of the ratio $\varrho$.

iii) Equation \eqref{AFC:c} shows that $\varphi_1(0)\,\varphi_1^\prime(+0) <0$.

Now it is not hard to obtain the asymptotic of the function $g_{tt}(r)$ for $r=GM y \to \infty$ in the form
\ben
g_{tt}(r) = 1-{\frac {2GM}{r}}+O_2(1/r)= 1+2\varphi_N(r)+O_2(1/r)\quad \text{when}\quad r \to \infty.
\la{AFgtt}
\een

Thus, we obtain two physically very important corollaries:

\begin{corollary}
The mass $M=\varrho M_0$ is precisely the Kepler mass of the massive point particle in the Newton gravitational potential $\varphi_N(r)=-{\frac {GM}{r}}$. Hence, one can measure the well-defined physical quantity $M$, studying the orbits of the test bodies at large enough distances around the point particle in the Newton approximation.
\end{corollary}

\begin{corollary}
In GR we have $0<M<M_0$ and the mass defect $M_0-M>0$ of a massive point in GR is positive. Since gravitational binding energy $(M_0-M)c^2>0$ of GR-mass-point at rest is a positive quantity, this object is gravitationally stable.
\end{corollary}

On the other hand, making use of Eqs.\eqref{varphi_1},\eqref{x-y}, \eqref{AFgtt} and \eqref{jphi1} one obtains from the first order term in the Taylor series expansion of $g_{tt}(x)$ around the point $x_\infty$
\begin{subequations} \la{varphi1prime_0:a,b}
\ben
\varphi_1^\prime(+0)e^{-\bar\varphi(+0)}&=&1, \la{varphi1prime_0:a}\\
\varphi_1^\prime(-0)e^{-\bar\varphi(-0)}&=&0. \la{varphi1prime_0:b}
\een
\end{subequations}

Hence, we have the final results for admissible gauge functions $\bar\varphi(x)$
\ben
\varphi_1(x)=
\left\{
\begin{aligned}
&\Phi(x)+ \ln \varrho:&  \quad x \in [0,x_{max}), \\
&\hskip 1.15truecm \ln \varrho:&    x<0,\hskip 1.1truecm
\end{aligned}
\right.
\la{varphi_1_final}
\een
\ben
\varphi_1^\prime(x)=
\left\{
\begin{aligned}
&e^{-\bar\varphi(x)}>0:& \quad  x \in [0,x_{max}), \hskip .5truecm \\
&\hskip .25truecm 0  \hskip 1.truecm :&  \hskip -1.truecm  x < 0,\hskip 1.6truecm
\end{aligned}
\right.
\la{varphi_1prime_final}
\een
and
\ben
g_{tt}(x)=
\left\{
\begin{aligned}
&\varrho^2 e^{2\Phi(x)}:&  \quad x \in [0,x_{max}), \\
&\varrho^2\hskip .75truecm :& \hskip 1.15truecm    x<0.\hskip 1.1truecm
\end{aligned}
\right.
\la{gtt_final}
\een

\subsection{Imposing the constraint \eqref{w}}\la{Constraint}
Finally, we are ready to impose the constraint \eqref{w}. After some simple algebra it gives the final results
for admissible gauge functions $\bar\varphi(x)$
\ben
C\stackrel{w}{=}\pm 1,\quad  \text{and}
\la{C}
\een
\ben
e^{\varphi_2(x)}\stackrel{w}{=}
\left\{
\begin{aligned}
{\frac {2 \varrho e^{\Phi(x)} }{1- \varrho^2e^{2\Phi(x)}}}& :\quad x \in [0,x_{max}), \\
{\frac {1}{\ln{\tfrac 1 \varrho}-\Phi(x)}}& : \quad   x<0.\hskip 1.1truecm
\end{aligned}
\right.
\la{varphi_2_final}
\een

\begin{corollary}
As a result of the final formulae \eqref{varphi_1_final} and \eqref{varphi_2_final} we obtain
\ben
\rho(x)\stackrel{w}{=}
\left\{
\begin{aligned}
{\frac {2 G M}{1- \varrho^2e^{2\Phi(x)}}}& :\quad x \in [0,x_{max}), \\
{\frac {GM} {\varrho\left( \ln{\tfrac 1 \varrho}-\Phi(x)\right)}}& : \quad   x<0,\hskip 1.1truecm
\end{aligned}
\right.
\la{rho_x_final}
\een
and the most important physical result: In the physical domain there exists a minimal value of the luminosity radius
\ben
\rho_{min}=\rho(0)\stackrel{w}{=}{\frac {2 G M}{1- \varrho^2}}>2 G M=\rho_g.
\la{rho_min}
\een
\end{corollary}

It is remarkable that the physical result described by Eq.\eqref{rho_min} 
is gauge invariant since it does not depend on the choice of the function $\bar\varphi$.

In Eq.\eqref{rho_min}, the quantity $\rho_g=2 G M$ is the well-known gravitational radius of the mass $M$, often called also "the Schwarzschild radius", or "event horizon radius" of the Schwarzschild Black Hole (SBH) \cite{books}).
\begin{corollary}
Since $\varrho \in (0,1)$, formula \eqref{rho_min} shows that even the most compact matter object, such as a massive point, can not reach its gravitational radius $\rho_g$.
\end{corollary}

As in the case of all other spherically symmetric matter bodies, the gravitational radius $\rho_g$ remains inside the nonphysical domain of the body, 
i.e. in the domain $\rho \in [0,\rho_{min})$, which corresponds to $x<0$. 
From a physical point of view, one can consider this domain as "an optical illusion", or a "mirage" \cite{Fiziev2019}.

The jump $j_{\rho}(0)$ of the luminosity radius $\rho(x)$ at the point $x=0$ depends in a quite nontrivial way on the mass ratio $\varrho$:
\ben
j_{\rho}(0)=\rho(+0)-\rho(-0)\stackrel{w}{=} G M_0 \left( {\frac {1}{1-\varrho}}-{\frac {1}{1+\varrho}} +{\frac {1}{\varrho\ln\varrho}}\right)<0
\quad \text{for} \quad \varrho \in (0,1).
\la{jump_rho_0}
\een

This quantity is also a gauge-independent measurable physical quantity.

For the component $g_{rr}(x)$ we obtain
\ben
g_{rr}(x)\stackrel{w}{=}
\left\{
\begin{aligned}
& {\frac {16 \varrho^2 e^{2\Phi(x)-2\bar\varphi(x)}} { \left(1- \varrho^2 e^{2\Phi(x)}\right)^4}}:&  \quad x \in [0,x_{max}), \\
& {\frac { e^{-2\bar\varphi(x)} }  {\varrho^2 \left(\ln { \frac 1 \varrho} -\Phi(x)\right)^4}  }
\hskip .truecm :& \hskip 1.15truecm    x<0.\hskip 1.1truecm
\end{aligned}
\right.
\la{grr_final}
\een

The jump $j_{g_{rr}}(0)$ of the component of metric $g_{rr}(x)$ at the point $x=0$ depends also 
in a quite nontrivial way on the mass ratio $\varrho$ and on the values $\bar\varphi(\pm0)$ of the gauge function at the same point:
\ben
j_{g_{rr}}(0)=g_{rr}(+0)-g_{rr}(-0)\stackrel{w}{=} {\frac 1 {\varrho^2}}\left( \left({\frac {e^{-\bar\varphi(+0)/2}} {{\frac 1 2}
\left({\frac {1}{\varrho}}-\varrho\right)}}\right)^4 -
\left({\frac {e^{-\bar\varphi(-0)/2}} {
\ln {\frac {1}{\varrho}}}}\right)^4\right)
\quad \text{for} \quad \varrho \in (0,1).
\la{jump_grr_0}
\een

The component of metric $g_{rr}(x)$ is continuous at the point $x=0$ when $j_{g_{rr}}(0)=0$. This gives
\ben
{\frac 1 2}\left({\frac {1}{\varrho}}-\varrho\right)=e^{\frac{1}{2}j_{\bar\varphi}(0)} \ln {\frac {1}{\varrho}}.
\la{continuos_grr}
\een
where $j_{\bar\varphi}(0)=\bar\varphi(+0)-\bar\varphi(-0)$ is the jump of the gauge function $\bar\varphi(x)$ at the point $x=0$.
Equation \eqref{continuos_grr} has a real solution $\varrho_{grr:continuous}$ if $j_{\bar\varphi}(0) > 0$.

However, even for $\varrho=\varrho_{grr:continuous}$ the metric \eqref{ds0} is not continuous
because of the nonzero jump \eqref{jump_rho_0} of the luminosity radius $\rho(x)$ at the point $x=0$.

\section{Solution of the Schwarzschild problem  in the Hilbert gauge}\la{SecHilbert}

The Hilbert choice of the radial variable $r=\rho=\sqrt{A/2\pi}$ is preferable, 
since the luminosity radius $\rho$ is a geometrical quantity that is coordinate independent.

As we already know from the previous Subsection \ref{Constraint}, 
the massive point particle is placed at the point $\rho_{min}$, see Eq.\eqref{rho_min}. 
Then, the physical domain of the point particle problem is the interval $\rho \in [\rho_{min},\infty)$.

In the case of a point particle, the interval $\rho \in [0,\rho_{min})$ may be considered as a non-physical one, 
i.e. as a kind of "optical illusion" or as a mirage.

\subsection{The Form of the Solutions in the Hilbert Gauge}\la{HilbertForm}

Using the results of the previous Subsections \ref{Solutions} and \ref{Constraint},
it is not hard to obtain the following basic form of the Schwarzschild-problem-solutions in the Hilbert gauge:
\ben
\Phi(\rho)\stackrel{w}{=}
\left\{
\begin{aligned}
& \ln \sqrt{1-{\frac {\rho_g}{\rho}}}+ \ln{\frac 1 \varrho}:&  \quad \rho \in [\rho_{min},\infty), \\
& \hskip .5truecm -{\frac {1}{2\varrho}}{\frac{\rho_g}{\rho}} + \ln{\frac 1 \varrho} :& \quad \rho \in [0, \rho_{min}),
\end{aligned}
\right.
\la{Phi_rho}
\een
\ben
j_{\Phi}(\rho_{min})={\frac 1 2}\left({\frac 1 \rho}-\rho\right)-\ln {\frac 1 \rho}.
\la{jPji_rho}
\een
\ben
g_{tt}(\rho)\stackrel{w}{=}
\left\{
\begin{aligned}
& 1-{\frac {\rho_g}{\rho}} :& \quad \rho \in [\rho_{min},\infty), \\
& \hskip .4truecm \varrho^2\hskip .3truecm:& \quad \rho \in [0, \rho_{min}),
\end{aligned}
\right.
\la{gtt_rho}
\een
\ben
j_{g_{tt}}(\rho_{min})=0.
\la{j_gtt_rho}
\een
\ben
g_{\rho\rho}(\rho)\stackrel{w}{=}
\left\{
\begin{aligned}
& - \left(1-{\frac {\rho_g}{\rho}}\right)^{-1} :& \quad \rho \in [\rho_{min},\infty), \\
& \hskip .7truecm -1 \hskip 1.05truecm:& \quad \rho \in [0, \rho_{min}).
\end{aligned}
\right.
\la{grhorho}
\een
\ben
j_{g_{\rho\rho}}(\rho_{min})=1-{\frac 1 {\varrho^2}} < 0.
\la{j_grhorho_rho}
\een

These solutions depend on the pair of parameters $\rho_g$ and $\varrho$ according to Eqs.\eqref{rho_min}, \eqref{gtt_rho} and \eqref{grhorho}.
Thus, we have a two-parameter family of solutions of the Schwarzschild point mass problem.
\begin{comment}\la{comment1}
In formulae \eqref{Phi_rho}, \eqref{gtt_rho}, and \eqref{grhorho}, we restrict our consideration to the domain $\rho \in [0,\infty)$,
where all used quantities are real and there is no change of the physical meaning of the variables $t$ and $\rho$.
\end{comment}
\begin{comment}\la{comment2}
It is easy to check directly that solutions \eqref{gtt_rho} and \eqref{grhorho} obey the Einstein equation for the Schwarzschild problem in the Hilbert gauge with the corresponding Dirac-$\delta$-function mass-term, see Eqs.(26) in \cite{Fiziev2019}.
\end{comment}
\begin{comment}\la{comment3}
In the course of deriving Eqs. \eqref{Phi_rho}, \eqref{gtt_rho}, and \eqref{grhorho} from Eqs. \eqref{varphi_1_final}, \eqref{gtt_final}, and \eqref{grr_final} and making use also of the relation $g_{\rho\rho}=g{rr}\left({\frac{d\rho}{dr}}\right)^{-2}$, one sees that the gauge function $\bar\varphi$ disappears because of the corresponding cancelations. This shows that  Eqs. \eqref{gtt_rho} and \eqref{grhorho} are rho-gauge invariant, as it should be, since the variable $\rho$ is a geometrical quantity, whose values do not depend on the choice of the radial variable.
\end{comment}
\begin{comment}\la{comment4}
One more advantage of the Hilbert form of the SMP solution is that only the $g_{\rho\rho}$
component of the metric is discontinuous at the place of the massive point source, see Eqs. \eqref{j_gtt_rho} and \eqref{j_grhorho_rho}.
Note also that in this gauge the $\rho$ variable is continuous by construction.
\end{comment}

\subsection{SMP in Physical Gauge}

One can easily overcome the unusual coordinate place $\rho=\rho_{min}$ of the massive point in the Hilbert gauge going to the new variable
$r = \rho-\rho_{min} \leftrightarrows \rho = r + \rho_{min}$. After this simple linear transformation we obtain from the relations
\ben
\Phi(r)\stackrel{w}{=}
\left\{
\begin{aligned}
& \ln \sqrt{\frac {r+\rho_g \varrho^2/\left(1-\varrho^2\right)}{r+\rho_g /\left(1-\varrho^2\right)}}+ \ln{\frac 1 \varrho}:&  \quad r \in [0,\infty), \\
& \hskip .5truecm -{\frac {1}{2}}{\frac{\rho_g/\varrho}{r+\rho_g /\left(1-\varrho^2\right)}} + \ln{\frac 1 \varrho} :& \quad r \in [-\rho_{min},0),
\end{aligned}
\right.
\la{Phi_r}
\een
\ben
g_{tt}(r)\stackrel{w}{=}
\left\{
\begin{aligned}
& {\frac {r+\rho_g \varrho^2/\left(1-\varrho^2\right)}{r+\rho_g /\left(1-\varrho^2\right)}}:& \quad r \in [0,\infty), \\
& \hskip .4truecm \varrho^2\hskip .3truecm:& \quad r \in [-\rho_{min},0),
\end{aligned}
\right.
\la{gtt_r}
\een
\ben
g_{rr}(\rho)\stackrel{w}{=}
\left\{
\begin{aligned}
& - {\frac {r+\rho_g /\left(1-\varrho^2\right)}{r+\rho_g \varrho^2/\left(1-\varrho^2\right)}} :& \quad r \in [0,\infty), \\
& \hskip .7truecm -1 \hskip 1.05truecm:& \quad r \in [-\rho_{min},0).
\end{aligned}
\right.
\la{grr_r}
\een

As we can see, the obtained SMP-solution allows penetration in the domain $[-\rho_{min},0)$ "behind" the position of the mass point at $r=0$.

In subsequent paper \cite{Fiziev2019c}, we show that because of the Dirac $\delta$-function in Eqs. \eqref{dfR}, \eqref{fdf} (See  Appendix \ref{Appendix1}.),
the presence of SMP in spacetime raises an infinite semi-penetrable $\delta$-barrier (mirror) for GWs. This barrier is placed at the point $r=0$, which is the same as $\rho=\rho_{min}$ - in terms of the luminosity radius.

The coefficients of reflection and transition of the semi-penetrable mirror depend on the mass ratio $\varrho$.

We dubbed this case the "SMP with a soft core".

One can forbid penetration behind the position of the mass point, replacing the SMP-soft-core with a hard one, i.e. replacing the
infinite semi-penetrable $\delta$-barrier (mirror) with a not penetrable infinite barrier at the point $r=0$. This is equivalent to the standard exclusion
of the domain $r<0$ in spherical coordinates.

In terms of the luminosity radius the last procedure puts a totally reflecting mirror at the point $\rho=\rho_{min}$.
It was introduced and studied in detail for the first time in \cite{Fiziev2006},
without paying attention to the appearance of echoes in spreading waves under the corresponding quite special conditions.

\subsection{The Geometry of Spacetime of a Massive Point Particle in GR}\la{Geometry}

The geometry of the spacetime in the physical domain outside the source $\rho \in (\rho_{min},\infty)$\footnote{For simplicity, in this subsection
 we continue to write down all formulas using the luminosity distance $\rho$ instead of a more physical variable $r=\rho - \rho_{min}$, introduced in the previous subsection}
  is described by the following
\begin{corollary}
As a result of Eqs.\eqref{gtt_rho} and \eqref{grhorho}, in the physical domain $\rho \in (\rho_{min},\infty)$
of the luminosity radius, the Birkhoff theorem \cite{books} is strictly valid for all solutions of the SMP problem, found in the present paper.
\end{corollary}

Hence, for $\rho \in (\rho_{min},\infty)$ it is impossible to find any deviations from well-known experimental
and observational tests of GR  \cite{books} for motion of test bodies in the gravitational field of massive point particles.

 Thus, outside the point particle, the geometry of spacetime is precisely the same as in the case of one-parametric family of vacuum static spherically symmetric solutions of the Einstein equation, often dubbed in the literature the "Schwarzschild solution".

 In the case of the original Schwarzschild point particle problem, the existence of minimal physical value 
 of the luminosity radius $\rho_{min}$, defined by Eq. \eqref{rho_min}, 
 yields a very unusual geometry when one approaches the very mater particle in the limit $\rho \to \rho_{min}+0$:

In the Hilbert gauge, for $\rho=\rho_{min}+\delta\rho$, $0<\delta\rho/\rho_g \ll 1$ one obtains from Eqs.\eqref{gtt_rho}, \eqref{grhorho} and \eqref{ds0}
\ben
ds^2 \stackrel{w}{=}\varrho^2 dt^2 -{\frac 1 {\varrho^2} } \delta\rho^2 -\left({\frac {\rho_g}{1- \varrho^2}}\right)^2 d\Omega^2.
\la{limit_ds0}
\een

Below we discuss the peculiar properties of this spacetime geometry in more detail.

Formula \eqref{ldistance} gives for the radial geometrical distance from the point particle
\ben
l(\rho)&\stackrel{w}{=}& \rho \sqrt{1-{\frac {\rho_g}{\rho}}} - \rho_{min} \sqrt{1-{\frac {\rho_g}{\rho_{min}}}}
+{\frac {\rho_g}{2}}\ln\left( {\frac {\rho\left(1+\sqrt{1-{\frac {\rho_g}{\rho}}}\right)^2}         {\rho_{min}\left(1+\sqrt{1-{\frac {\rho_g}{\rho_{min}}}}\right)^2}} \right)=\nonumber\\ &=&{\frac{\delta\rho}{\sqrt{\varrho}}}+\mathcal{O}_2(\delta\rho) \quad \text{for}\quad \rho=\rho_{min}+\delta\rho, \quad 0<\delta\rho/\rho_g \ll 1.
\la{l_rho_distance}
\een

Note that in the case of a massive point particle in the Euclidean 3D space, when one approaches this particle, the geometrical distance goes to zero independently of particle mass.

In GR, the geometrical distance also goes to zero, when one approaches the point particle. However, in the case of GR this depends on the mass ratio $\varrho$.

Formula \eqref{Area} gives for the area of the matter-point-surface a {\em finite} value
\ben
A_{min}=A(\rho_{min})\stackrel{w}{=}{\frac {4\pi\rho_g^2}{\left(1-\varrho^2\right)^2}},
\la{Area_min}
\een
and
\ben
A_{\rho}=A(\rho_{min}+\delta\rho)\stackrel{w}{=}{\frac {4\pi\rho_g^2}{\left(1-\varrho^2\right)^2}}
+{\frac {4\pi\rho_g}{\left(1-\varrho^2\right)}}\delta\rho +\mathcal{O}_2(\delta\rho)\quad \text{for}\quad \rho=\rho_{min}+\delta\rho, \quad 0<\delta\rho/\rho_g \ll 1.
\la{Area_min+}
\een

Hence, when one approaches the massive point source, i.e. when according to Eq.\eqref{l_rho_distance} $l\to 0$, 
the luminosity of the massive point in GR remains finite, together with its surface $A_{min}>0$, see Eq.\eqref{Area_min}.
This is in a sharp contrast with the case of point particle in the Euclidean space where $A_{min}=0$.

Formula \eqref{lcircle} gives for the length of a big circle on the finite point mass surface
\ben
l_{c,min}=l_c(\rho_{min})\stackrel{w}{=}{\frac {2\pi\rho_g}{\left(1-\varrho^2\right)}}, \qquad l_c(\rho_{min}+\delta\rho)\stackrel{w}{=} l_{c,min}
+2\pi\delta\rho \quad \text{for}\quad \rho=\rho_{min}+\delta\rho>\rho_{min}.
\la{lc_min}
\een

Thus, in sharp contrast with the case of a massive point particle in the Euclidean space, 
in GR the length of a big circle on the finite surface of the point particle is also finite.

At the end, formula \eqref{3DV} gives for the 3D geometrical volume ${}^3V(\rho)$ of a sphere with the luminosity radius $\rho$, centered at the matter point
\ben
{}^3V(\rho)&\stackrel{w}{=}&
{\frac 4 3} \pi\left( {\frac {(\rho-\rho_g)(\rho-{\frac{3}4}\rho_g)(\rho+{\frac{1}2}\rho_g)}{\sqrt{1-{\frac{\rho_g}{\rho}}}}} -{\frac {(\rho_{min}-\rho_g)(\rho_{min}-{\frac{3}4}\rho_g)(\rho_{min}+{\frac{1}2}\rho_g)}{\sqrt{1-{\frac{\rho_g}{\rho_{min}}}}}} \right)-\nonumber\\
&-&{\frac \pi 4}\rho_g^3\ln\left({\frac{\rho}{\rho_{min}}}\left({\frac{1-\sqrt{1-{\frac{\rho_g}{\rho}}}}
{1-\sqrt{1-{\frac{\rho_g}{\rho_{min}}}}}}\right)^2\right)=\nonumber\\
&=&{\frac \pi 2} {\frac {(1+\varrho^2)\left( 5-(2-\varrho^2)^2\right)}{\varrho(1-\varrho^2)^2}}\rho_g^2 \delta\rho
+\mathcal{O}_2(\delta\rho) \quad \text{for}\quad \rho=\rho_{min}+\delta\rho, \quad 0<\delta\rho/\rho_g \ll 1.
\la{3DV_Hilbert}
\een

As seen, in GR the 3D geometrical volume ${}^3 V(\rho)$ of a sphere with the luminosity radius $\rho$ goes to zero when the point particle is approached, i.e. when according to Eq.\eqref{l_rho_distance} $l(\rho)\to 0$. In contrast to the Euclidean case, where ${}^3 V(\rho)\sim l(\rho)^3$ when $l(\rho)\to 0$, in GR this volume tends to zero linearly with the geometrical distance ${}^3 V(\rho)\sim l(\rho)$ when $l(\rho)\to 0$, not cubically.

Finally, in the domain $\rho \in [0,\rho_{min})$ the spacetime is flat, since the Riemann tensor vanishes identically: $R_{\mu,\nu,\lambda}^\kappa \equiv 0$.
This is consistent with the GR solution for vacuum inside a spherically symmetric shell, as well as with the Newton gravity,
according to which there is no gravitational force inside such shell.

In the case of SMP-solution there is no real physical shell
but just domain $\rho \in [0,\rho_{min})$ "behind" mass point,
which resembles a shell, when considered in terms of the luminosity radius $\rho$.

In this domain we have the unusual squared 4D interval:
\ben
ds^2=\varrho^2 dt^2 - d\rho^2-\rho^2 d\Omega^2.
\la{ds2_0}
\een

\section{Concluding Remarks}\la{Conclusion}

In the present paper, for the first time, we present a correct solution of the Schwarzschild Massive Point (SMP) problem in arbitrary radial gauge,
see Sections \ref{Gauges}, \ref{SPS},
and discuss in detail the significant difference between the SMP solution and the vacuum the Schwarzschild solution of the Einstein equations
\eqref{Einst} called the Schwarzschild Black Holes (SBH).

We use the same standard spherical coordinates in the flat 3D Euclidean tangent space $T{}^3\!\mathcal{E}$ above the place of the SMP,
and in the 3D Riemann Manifold ${}^3\!\mathcal{M}$ of the SMP, see Section  \ref{3DRG}.
The origin of the corresponding coordinate systems in both spaces is chosen
at the place of the massive point particle under consideration.

Using normal field variables $\varphi_1$, $\varphi_2$ and $\bar\varphi$, in Section \ref{NormalCoordinates}
we found an exact general solution of the system of ordinary differential
equations for the metric of the SMP in arbitrary radial gauge, see Section \ref{GeneralSolution},
obtained by a proper variational principle in Section \ref{VP},
and impose the corresponding additional conditions, see Sections \ref{Solution_b} and \ref{Constraint}.

Special attention is paid to the formulation of strict mathematical assumptions needed for a correct solution of the SMP problem.

The solution for the SMP has an unavoidable jump of the metric at the place of the SMP
needed to comply with the Einstein equations with the distributional energy-momentum tensor of type \eqref{EM_MPs}.

This way, we obtained a two-parameter family of solutions of the SMP, in sharp contrast with the one-parameter family of SBH solutions.

The two parameters of SMP solutions are the gravitational radius $\varrho_g = 2 G M/c^2 \in (0,\infty)$ and mass ratio $\varrho = M/M_0 \in (0,1)$,
where $M>0$ is the Kepler mass, and $M_0>M$ is the bare mass of SMP.

The most important physical result is the discovery of minimal value $\rho_{min}=\rho_g/\left(1-\varrho^2\right)>\rho_g$ of the luminosity radius of SMP.
It shows that the luminosity of a mass point tends to a finite value when one approaches this point.

A bigger mass ratio corresponds to darker SMPs and to a smaller mass-defect $M_0-M$.

The unusual geometry around the SMP is considered in Section \ref{Geometry}.
As we showed, in GR the geometry around a mass point is very different from the geometry around empty points.

Physically, this is completely natural, since in the case of SMP we have infinite density of finite mass concentrated in zero volume
at the place of the very point. Such density is described by the Dirac $\delta$-function.
Then, the Einstein equations with such source of mass define infinite curvature at the place of a massive point and other unusual
properties of geometry, which are considered in detail in Section \ref{Geometry}.

From a physical point of view, it is clear that being the most compact object in GR,
the SMP can be considered as a spherically symmetric static body of finite dimension $R$,
as seen by an observer from the large 3D distance $l\gg R$.

The condition $\rho_{min} \geq R$, where $R$ is the physical radius of a spherically symmetric body 
with the Kepler mass $M$, will offer a possibility to observe the effect
of jump of geometry at the point $\rho_{min}$, see Section \ref{Geometry}. This condition gives $\varrho \geq \sqrt{1-\rho_g/R}$ and
$$ E_{binding}= \left(M_0-M\right)c^2 \leq Mc^2 \left({\frac{1}{\sqrt{1-\rho_g/R}}}-1\right) = Mc^2 \left( {\frac {\rho_g}{2R}}+\mathcal{O}_2\left({\frac {\rho_g}{2R}}\right) \right).$$

For fixed Kepler mass $M$, in the limit $R \to \infty$ we obtain $\varrho \to 1-0$ and $E_{binding} \to +0$, i.e., the body becomes gravitationally unstable.

This consideration gives some idea why we may not see bodies with large
$\rho_{min} \gg \rho_g \Rightarrow 0 < 1-\varrho^2 \ll 1$.
Such dark bodies will not be stable gravitationally and can be easily destroyed, for example, by slow rotation.

The limiting case $\varrho=1$ indicates the absence of mass defect and is not admissible in GR.
This would be a non-physical absolutely dark and infinitely large object with $\rho_{min}=\infty$ and zero binding energy $E_{binding}$.
Gravitationally, this type of object would be absolutely not stable and would be destroyed by any infinitely small perturbation.

This conclusion needs careful study of more realistic spherically symmetric bodies of finite dimension and some realistic matter-equation-of-state.
This problem is beyond the scope of the present paper.

The opposite case of ECOs is intensively discussed in the literature, see the review paper \cite{Cardoso2019} and a large number of references therein.

The ECOs$\neq$BHs were introduced as alternatives to BHs, especially, to study possible echoes in observed GW during the ringing phase of binary merger.
For this purpose, the ECOs$\neq$BHs need to have a luminosity radius well-below the luminosity radius of the photon sphere $\rho_{ps}$, i.e.,
$\rho_{min}<\rho_{ps}={\frac 3 2}\rho_G$.

For appearance of observable echoes, the condition $\rho_{min}=\rho_g (1+\varrho^2)$ with $0<\varrho^2 \ll 1$ is preferable \cite{Fiziev2019}.
Then, we obtain $M_0\sim M/\varrho \gg M$.
Hence, the binding energy $\left(M_0-M\right)c^2 \sim M c^2 (1/\varrho-1) \gg Mc^2$.
Such ECOs are strongly bounded by gravity and are very stable.

When the Keplerian mass $M>0$ is fixed,  for $\varrho \to 0$ we obtain a limiting case with
$\rho_{min}=\rho_G+0$,  $\varrho =+0$,  infinite $M_0=1/(+0)$ and infinite positive binding energy.
This is not a BH, since at $\rho=\rho_{min}$ we have a reflecting mirror described by the corresponding
jump of the metric at the point $\rho=\rho_{min}$, see Appendix \ref{Appendix1} and \cite{Fiziev2019c} for a detailed consideration.

In Appendix \ref{Appendix1}, we have developed the theory of regularization of a specific class of distributions, which appear in the SMP problem.

The main idea is that the value of the kernel of the linear functional, which defines distribution on a proper class of test functions, is not essential
and can be changed in a convenient way.

The only such kernels which appear, when solving the SMP are usual functions with a finite jump at the place of the SMP.
The natural and unambiguous regularization of them is to prescribe to such functions their average value at the point of the jump.
This property is borrowed from the continuous function and does not change the distribution.

This regularization procedure allows us to deliver many of the properties of continuously differentiable functions to this specific
class of distributions and to use them in solution of the SMP problem, see Appendix \ref{Appendix1}.

Of course, such regularization does not solve analogous problems for a more wide
class of distributions \cite{Gelfand}, but we do not need this for solution of the SMP problem.

\appendix

\section{Distributions and regularization of discontinuous functions}\la{Appendix1}

Consider a function $f(x)$ defined on the real axis $(-\infty,\infty)$, which is discontinuous at the point $x=0$.
Let $f_{+}(x)\in \mathcal{C}^\infty_{[0,\infty)}$ and $f_{-}(x)\in \mathcal{C}^\infty_{(-\infty,0]}$ be infinite differentiable functions
defined in the corresponding intervals $[0,\infty)$ and $(-\infty,0]$. The derivatives at the point $x=0$ of $f_{+}(x)$ and $f_{-}(x)$
are defined as right and left derivatives, respectively.
The limits $f_{+}(+0)$ and $f_{-}(-0)$ do exist.

Then, there exist three different functions discontinuous at the point $x=0$:
\begin{subequations} \la{f:abc}
\ben
f_a(x):=
\left\{
\begin{aligned}
& f_{+}(x):& \quad x>0, \la{fa}\\
& f_{-}(x):& \quad x<0.
\end{aligned}
\right.  \\
f_b(x):=
\left\{
\begin{aligned}
& f_{+}(x):& \quad x>0, \la{fb}\\
& f_{-}(x):& \quad x\leq 0.
\end{aligned}
\right.  \\
f_c(x):=
\left\{
\begin{aligned}
& f_{+}(x):& \quad x \geq 0, \la{fc}\\
& f_{-}(x):& \quad x<0.
\end{aligned}
\right.
\een
\end{subequations}

All three functions have equal jumps at the point $x=0$
\ben
j_{f_a}(0)=j_{f_b}(0)=j_{f_c}(0):=f_{+}(+0)-f_{-}(-0).
\la{jf}
\een
and equal average values
\ben
<f_a>_{0}=<f_b>_{0}=<f_c>_{0}:={\frac 1 2}\left(f_{+}(+0)+f_{-}(-0)\right).
\la{jf}
\een
When considered as kernels of regular linear functionals, i.e. as distributions \cite{Gelfand}, the three functions $f_{a,b,c}$ define the same distribution
\ben
(f_a,\psi)=(f_b,\psi)=(f_c,\psi):=\int_{-\infty}^0 f_{-}(x)\psi(x)dx+\int_{0}^{\infty} f_{+}(x)\psi(x)dx
\la{distr_f}
\een
on linear space of the corresponding test functions $\psi(x) \in \mathcal{C}^\infty_{(-\infty,\infty)}$.
Proper boundary conditions at $x=\pm \infty$ are supposed to guarantee the convergence
of the integrals for all admissible test functions\footnote{The symbol $\stackrel{d}{=}$ denotes equality in the sense of distributions.
Thus, in this Appendix we have two types of equalities: equalities of functions, and equalities of distributions.}.

However, none of the functions $f_{a,b,c}(x)$ obeys the relation $f(0)=<f>_{0}$,
needed to make transparent some important properties of the specific distributions,
which appear in our consideration.

Note that the relation $f(0)=<f>_{0}$ is certainly valid for  functions that are continuous at the point $x=0$.

Therefore, we define  the {\em regularized} discontinuous function $f_R(x)$ with the property $f_R(0)=<f_R>_{0}$ in the following way:
\ben
f_{R}(x)=
\left\{
\begin{aligned}
 f_{+}(x): \quad x>0,  \\
 <f>_0:  \quad x=0, \\
 f_{-}(x): \quad x<0.
\end{aligned}
\right.
\la{fR}
\een

It is obvious that the regularized function $f_R(x)$ defines the same regularized linear functional \eqref{distr_f},
i.e. the same distribution, since this distribution does not depend on the value of its kernel at the point $x=0$.
Hence, all discontinuous functions $f_{a,b,c}$ and $f_R(x)$ are equivalent in the sense of kernels of distributions, i.e.,
$f_a\stackrel{d}{=}f_b\stackrel{d}{=}f_c\stackrel{d}{=}f_R$.
However, the regularized function is unique and defines distribution $(f_R,\psi)$ with distinguished properties, as shown below.

In particular, consider the Heaviside $\Theta(x)$ function. Following the book by I.~M.~Gel'fand and G.~E.~Shilov in reference \cite{Gelfand},
\ben
\Theta(x):=
\left\{
\begin{aligned}
& 1:& \quad x>0, \la{Theta}\\
& 0:& \quad x<0.
\end{aligned}
\right.
\een

It defines the distribution $(\Theta,\psi)=\int_0^\infty \psi(x)$. 
The derivative $\Theta^\prime=\delta(x)$ defines the Dirac $\delta$-function, i.e. the distribution
$(\delta,\psi)=\psi(0)$.

The corresponding regularized $\Theta_R(x)$ function
\ben
\Theta_R(x):=
\left\{
\begin{aligned}
 1: \quad x>0,  \\
 1/2:  \quad x=0, \\
 0: \quad x<0.
\end{aligned}
\right.
\een
is the kernel of regular linear functional that defines the same distribution as function \eqref{Theta} and produces two functions $\Theta_{\pm}(x)=\Theta_R(\pm x)$
\begin{subequations} \la{Th:ab}
\ben
\Theta_{+}(x):=
\left\{
\begin{aligned}
 1: \quad x>0,  \\
 1/2:  \quad x=0, \\
 0: \quad x<0.
\end{aligned}
\right.  \\
\Theta_{-}(x):=
\left\{
\begin{aligned}
 0: \quad x>0,  \\
 1/2:  \quad x=0, \\
 1: \quad x<0.
\end{aligned}
\right.
\een
\end{subequations}
with the following basic properties:

i) $\Theta_{\pm}(x)=\Theta_{\mp}(-x)$.

ii) $(\Theta_{+},\psi)=\int_0^\infty \psi(x)dx$ and $(\Theta_{-},\psi)=\int_{-\infty}^0 \psi(x)dx$.

iii) $\Theta_{+}^\prime(x)=\delta(x)$ and $\Theta_{-}^\prime(x)=-\delta(x)$.

iv) $j_{\Theta_\pm}(0)=\pm 1$ and $\Theta_\pm(0) = <\Theta_\pm>_0=1/2$.

v) $\Theta_{+}(x) + \Theta_{-}(x) = 1$.

vi) $\Theta_{+}(x)  \Theta_{-}(x)\stackrel{d}{=} 0$.

vii) $\Theta_{+}(x)\stackrel{d}{=}\Theta_{+}(x)^2\stackrel{d}{=}\Theta_{+}(x)^3=\dots$

viii) The functions $\Theta_\pm(x)$ define a basis for representation of the regularized discontinued functions in the form
\ben
f_R(x)=f_{+}(x)\Theta_{+}(x)+ f_{-}(x)\Theta_{-}(x).
\la{fR_Theta}
\een

Some consequences can be easily derived from the above basic properties:

1. If $F(x)$ is a real analytic function in some vicinity of the point $x=0$ that has the Taylor series expansion
$F(x)=\sum_{n=0}^\infty {\frac {1}{n!}} F^{(n)}(0) x^n$, then using the properties v) and vii) it is not hard to obtain
$$ F\left(x \Theta_\pm(x)\right))\stackrel{d}{=}F(x)\Theta_\pm(x)+F(0)\left(1-\Theta_\pm(x)\right).$$

In particular,
$$\exp \left(x \Theta_\pm(x)\right)=\exp(x)\Theta_\pm(x) +1-\Theta_\pm(x).$$

We need this result in the basic text, for example, in calculation of $g_{tt}$, using the solution of the SMP problem $\varphi_1(r)$.

2. From the properties vi) and vii) one obtains

$$f_R(x)^n \stackrel{d}{=} f_{+}(x)^n\Theta_{+}(x)+f_{-}(x)^2\Theta_{-}(x), \quad n=1,2,3\dots$$

3. From the properties iii) and viii) one obtains
\ben
f_R(x)^\prime= f_{+}(x)^\prime\Theta_{+}(x)+f_{-}(x)^\prime\Theta_{-}(x)+ j_{f_R}(0)\delta(x).
\la{dfR}
\een
and
\ben
\left(f_R(x)^2\right)^\prime \stackrel{d}{=} 2 f_R(x)f_R(x)^\prime.
\la{df2_2fdf}
\een

For example, if we apply relations \eqref{dfR} and \eqref{df2_2fdf} for calculation of the expression
$f_{R}(\rho){\frac {df_{R}(\rho)}{d\rho}}$ with
$$f_{R}(\rho)=\sqrt{\frac {g_{tt}(\rho)}{-g_{\rho\rho}(\rho)}}$$
defined by the metric of the SMP \eqref{gtt_rho},\eqref{grhorho} in the Hilbert gauge, the result is:
\ben
f_{R}(\rho){\frac {df_{R}(\rho)}{d\rho}}\stackrel{d}{=}\left(1-\rho_g/\rho\right)\Theta_{+}(\rho-\rho_{min}) - {\frac 1 2}(1-\varrho)\varrho(1+\varrho)\delta(\rho-\rho_{min})
\la{fdf}
\een
This formula plays an essential role in our paper \cite{Fiziev2019c}.

In the case of distributions of a more general type, we have no such specific properties.
In particular, multiplication of distributions is not well defined in the general case.

Even in our case of a very special class of distributions, there is no generalization of relation \eqref{df2_2fdf} of type
\ben
\left(f_R(x)^n\right)^\prime \stackrel{d}{=} n f_R(x)^{n-1}f_R(x)^\prime
\la{dfn_nfdf}
\een
for degrees $n>2$ without some additional algebraic conditions on $f_{+}(+ 0)$ and $f_{-}(-0)$.

For example, relation \eqref{dfn_nfdf} is valid if $f_{+}(+ 0) f_{-}(-0)=0$, 
which is the case of themetric of the SMP \eqref{gtt_rho},\eqref{grhorho} in the Hilbert gauge.
Fortunately, we do not need such generalizations for arbitrary $f_{+}(+ 0)$ and $f_{-}(-0)$.

For two regular distributions with the kernels $f_R(x)$ and $h_R(x)$ with discontinuity at the same point $x=0$ one easily obtains from the above basic rules more general relations
$$h_R(x) f_R(x) \stackrel{d}{=} h_{+}(x) f_{+}(x)\Theta_{+}(x)+h_{-}(x) f_{-}(x)\Theta_{-}(x)$$
and
$$\left(h_R(x) f_R(x)\right)^\prime \stackrel{d}{=} \left(h_{+}(x) f_{+}(x)\right)^\prime\Theta_{+}(x)+\left(h_{-}(x) f_{-}(x)\right)^\prime\Theta_{-}(x)
+\big(<h_R>_0 j_{f_R}(0)+<f_R>_0 j_{h_R}(0)\big)\delta(x).$$

It is clear that making a simple translation $x \mapsto x+x_0$, we can obtain
analogous properties of the functions discontinued at an arbitrary point $x=x_0 \in (-\infty,\infty)$.

The properties of the functions that have discontinuity at some countable number of points $x_1,x_2,x_3,\dots$
can be obtained using the corresponding superposition of the above rules separately for each of these points.

The above properties can be derived also in the framework of the sequential approach to the distributions.

As seen, due to our definition \eqref{fR}, the regularized discontinuous function $f_R(x)$
defines distributions with some important properties,
analogous to the properties of the continuous differentiable functions.

These properties allow us to use such special class of distributions
in solving the Einstein equations \eqref{Einst} for the SMP and energy-momentum tensor of distributional type \eqref{EM_MPs}.


\begin{thebibliography}{}
\bibitem{Schwarzschild} K.~Schwarzschild, Sitzungsber. Preus. Akad. Wiss. Phys. Math. Kl., p. 189 (1916).
%
\bibitem{Hilbert1917} D.~Hilbert, Nachr. Ges. Wiss. G\"otingen, Math. Phys. Kl., 53 (1917).
%
\bibitem{books} J. L. Synge, {\em Relativity: The General Theory}, North Holland Publ. Comp., Amsterdam (1960).

                  L.~D.~Landau, E.~M.~Lifshitz, {\em The Classical Theory of
                  Fields}, 2d ed.; Reading, Mass: Addison-Wesley, 1962;

                  S.~Weinberg,{\em Gravitation and Cosmology}, Wiley,
                  N.Y., 1972;

                  C.~Misner, K.~S.~Thorne, J.~A.~Wheeler, {\em Gravity},
                  W.~H.~Freemand\&Co.,1973.

                  S. Chandrasekhar, The Mathematical Theory of Black Holes, Oxford University Press, New York (1983).

                  R. M. Wald, {\em General Relativity}, The University of Chikago Press, Chikago and London, 1984.

                  V.~P.~Frolov, I.~D.~Novikov, {\em Black Hole Physics}, Kluwer Acad. Publ., 1998.

\bibitem{EWeynberg} E. J. Weinberg, {\em Magnetically Charged Black Holes with Hair}, Lectures presented at the XIII International Symposium “Field Theory and
Mathematical Physics”, Mt. Sorak, Korea (June-July 1994). arXiv:gr-qc/9503032.
%
\bibitem{LOGOOPEN} https://losc.ligo.org
%
%
\bibitem{Barack2018}Leor Barack et al, {\em Black holes, gravitational waves and fundamental
physics: a roadmap}, arXiv:1806.05195.

\bibitem{RW} Regge T and Wheeler J A, Phys. Rev. 108 1063 (1957).
%
\bibitem{Starobinskij1973} A. A. Starobinskij and S. M. Churilov, Zhurnal Eksperimentalnoi i Teoreticheskoi Fiziki {\bf 65} 3 (1973).
%

\bibitem{Z} Zerilli F J, Phys. Rev. D 9 860 (1974).
%
\bibitem{ChandraDetweiler1975} S. Chandrasekhar and S. L. Detweiler, Proc. Roy. Soc. Lond. A {\bf 344}  441-452 (1975).
%
\bibitem{Kokkotas1999} K.D. Kokkotas, B.G. Schmidt, LivingRew.Rel.2:2(1999).
%
\bibitem{Nagar05}A. Nagar, L, Rezzolla,  Class. Quantum Grav. {\bf 22}  R167 (2005).
%
\bibitem{Fiziev2006} Plamen P Fiziev, CQG. {\bf 23}, 2447–2468 (2006).
%
\bibitem{Berti06} E. Berti, V. Cardoso, C. M. Will, PRD {\bf 73} 064030 (2006).
%
\bibitem{Berti09} E. Berti, V. Cardoso, and A. O. Starinets, Class. Quantum Grav. {\bf 26}  163001 (2009).
%
\bibitem{Fiziev2011} Plamen Fiziev, Denitsa Staicova, PRD {\bf 84}, 127502 (2011).
%
\bibitem{Nigel2016} Nigel T. Bishop · Luciano Rezzolla, {\em Extraction of gravitational waves in numerical relativity} Living Rev Relativ 19:2 (2016).
%
\bibitem{Testing2019} Maximiliano Isi, Matthew Giesler, Will M. Farr, Mark A. Scheel, Saul A. Teukolsky
     {\em  Testing the no-hair theorem with GW150914}, PRL {\bf 123}, 111102 (2019); arXiv:1905.00869.
%


%
\bibitem{Cardoso16} V. Cardoso, E. Franzin, P. Pani, Phys. Rev. Lett. 116, 171101 (2016).
%
\bibitem{Mark2017} Zachary Mark, Aaron Zimmerman, Song Ming Du, Yanbei Chen,{\em A recipe for echoes from exotic compact objects}, arXiv:1706.06155.
%
\bibitem{Cardoso2017} V. Cardoso, P. Pani,  {\em The observational evidence for horizons: from echoes to precision gravitational-wave physics}, arXiv:1707.03021.
%
\bibitem{Conklin18}Randy S. Conklin, Bob Holdom, and Jing Ren, {\em Gravitational wave echoes through new windows}, arXiv:1712.06517.
%
\bibitem{Wang2018} Qingwen Wang, Niayesh Afshordiy, {\em Black Hole Echology: The Observer's Manual}, arXiv:1803.02845.
%
\bibitem{Raposo18} G. Raposo, P. Pani1, M. Bezares, C. Palenzuela, V. Cardoso,
             {\em Anisotropic stars as ultracompact objects in General Relativity}, arXiv:1811.07917 (2018).
%
\bibitem{Cardoso2019} V. Cardoso, P. Pani, {Testing the nature of dark compact objects: a status report}, Living Reviews in Relativity (2019); https://doi.org/10.1007/s41114-019-0020-4.
%

\bibitem{EHT} The Event Horizon Telescope Collaboration,
%
The Astrophysical Journal Letters, 875:L1 (2019);
%
The Astrophysical Journal Letters, 875:L4 (2019);
%
The Astrophysical Journal Letters, 875:L5 (2019);
%
The Astrophysical Journal Letters, 875:L6 (2019).
\bibitem{Maggiore08} M. Maggiore {\em Gravitational Waves}, Oxford University Press, 2008.
%
\bibitem{Jafari19} A. Jafari, {\em GRAVITATIONAL RADIATION FROM BINARIES: A PEDAGOGICAL INTRODUCTION}, arXiv:1908.04410.
%

\bibitem{Fiziev2019} Plamen P. Fiziev {\em The Era of Gravitational Astronomy and Gravitational Field of Non-Rotating Single Point Particle in General Relativity}, ISSN 1063-7796, Physics of Particles and Nuclei, 2019, Vol. 50, No. 6, pp. 944–972. © Pleiades Publishing, Ltd., 2019.
%
\bibitem{GH}  R.~Gautreau, B.~Hoffmann, Phys. Rev. D{\bf 17}, 2552 (1978).
%
\bibitem{Eddington}  A.~S.~Eddington, {\em The mathematical theory of
relativity}, 2nd ed. Cambridge, University Press, 1930 (repr.1963).
%
%
\bibitem{Gelfand}   L.~Schwartz, {\em Th\'eorie des distributions}
                    I, II, Paris, 1950-51;\,
                    I.~M.~Gel'fand, G.~E.~Shilov, {\em Generalized
                    Finctions}, N.Y., Academic Press, 1964;\,
                    H.~Bremermann, {\em Distrinutions, Complex Variables and Fourier
                    Transform}, Addison-Wesley Publ. Co. Reading,
                    Massachusetts, 1965.
%
\bibitem{YB}Y.~Bruhat, {Cauchy Problem}, in {\em Gravitation: an
introduction to current research}, ed. L. Witten, John Wiley, N.Y. , 1962.
%
\bibitem{GT} R.~Geroch, J.~Traschen, Phys. Rev. D{\bf 36} 1017 (1987).
%
\bibitem{Exact}     D.~Kramer, H.~Stephani, M.~Maccallum,
                    E.~Herlt, Ed.~E.~Schmutzer, {\em Exact Solutions of the Einstein
                    Equations}, Deutscher Verlag der
                    Wissenschaften, Berlin, 1980.

                    H.~Stephani, D.~Kramer, M.~Maccallum,
                    E.~Herlt, {\em Exact Solutions of Einstein's
                    Field Equations}, Sec. Ed., Cambridge University
                    Press, 2003.
%
%
\bibitem{Dirac_Bargman} P. A. M. Dirac, Can. J. Math. {\bf 2} (2), 129-148 (1950).

 J. L. Anderson, P. J. Bergmann, Phys. Rev. {\bf 83} 1018 (1951).

 P. A. M. Dirac, Can. J. Math. {\bf 2} (2), 129-148 (1952).
%
\bibitem{Einstein1915}  Albert Einstein, Zirzungsber. Peuss. Akad. Wiss. {\bf 47}(2)pp. 831-839 (1915).
%
\bibitem{Einstein1916}  Albert Einstein, Annalen der Physik, Vierte Folge, Band {\bf 49} pp. 769-822 (1916).
%
\bibitem{Fiziev2019c} Plamen P. Fiziev {\em Exact Solutions to the Regge-Wheeler and Zerilli  Equations, Quasi Normal Modes and Echoes of Gravitational Waves}, in preparation.
%

\end{thebibliography}
\end{document}